\documentclass[12pt]{article}
\usepackage{epsf}
\title{ {\bf
The noncommutative effects on the dipole moments of fermions in
the standard model}}
\author{\vspace{1cm}\\
        {\bf E. O. Iltan}
        \thanks{E-mail address:
        eiltan@heraklit.physics.metu.edu.tr}
 \\
        Physics Department, Middle East Technical University \\
        Ankara, Turkey\\}

\date{}

\begin{document}
\setlength{\baselineskip}{24pt}
\maketitle
\setlength{\baselineskip}{7mm}
\begin{abstract}
We study the dipole moments, electric dipole moment, weak electric
dipole moment, anomalous magnetic moment, anomalous weak magnetic
moment, of fermions in the noncommutative extension of the SM. We
observe that the noncommutative effects are among the possible
candidates to explain the electric and weak electric dipole moment
of fermions. Furthermore, the upper bounds for the parameters
which carry space-time and space-space noncommutativity can be
obtained by using the theoretical and experimental results of the
fermion dipole moments.
\end{abstract}
\thispagestyle{empty}
\newpage
\setcounter{page}{1}
\section{Introduction}
The dipole moments of fermions arising from the couplings with the
photon and the Z boson provide precise tests of the quantum field
theories. In the case of the photon-fermion-fermion ($\gamma
\,f^+\, f^-$) coupling there exist two types of dipole moments:
the electric dipole moment (EDM) and the anomalous magnetic moment
(AMM). On the other hand, the weak dipole moments of the fermions
arise from the couplings of $Z f^+f^-$ vertex and they are two
types similar to previous ones: the weak electric dipole moment
(WEDM) and the weak anomalous magnetic moment (WAMM). Within the
SM the AMM (WAMM) receives its leading value from one-loop
corrections, however the EDM (WEDM) can exits at the higher order
in the coupling constant since it receives the non-vanishing value
through the CP violating effects coming from the CKM matrix
elements. Therefore the SM model contribution to EDM and WEDM is
irrelevant in the phenomenological analysis and forces one to
search the new physics effects, whereas the AMM and WAMM receive
the main contribution within the SM.

There are various studies on the the dipole moments of fermions
(EDM, WEDM, AMM, WAMM) in the literature
\cite{Commins}-\cite{Riad}. The EDM of fermions are interesting
from the experimental point of view since there are improvements
in the experimental limits of charged lepton EDM. EDM of electron,
muon and tau have been measured and the present limits are $d_e
=(1.8\pm 1.2\pm 1.0) \times 10^{-27}$ e-cm \cite{Commins},
$d_{\mu} =(3.7\pm 3.4)\times 10^{-19}$ e-cm \cite{Bailey} and
$d_{\tau} =3.1\times10^{-16}$ e-cm \cite{Groom}. The measurement
of the electron EDM  has been made using heavy atoms and the first
result is obtained as $d_e=-0.2\pm 3.2 \times 10^{-26}$ e-cm
\cite{Hudson}. The search for the EDM of tau using the reaction
$e^+ e^- \rightarrow \tau^+ \tau^-$ has been done and the
numerical values $Re [d_\tau]=(1.15\pm 1.7 )\times 10^{-17}$ e-cm,
$Im [d_\tau]=(-0.83\pm 0.86) \times 10^{-17}$ e-cm were obtained
in \cite{Inami}. In \cite{Barr}, it is emphasized that the
dominant contribution to the EDM of lighter leptons comes from the
two loop diagrams that involve one power of the Higgs Yukawa
couplings. In this work, the CP-violation is assumed to be
mediated by neutral Higgs scalars and the EDM of electron is
predicted at the order of the magnitude of $10^{-26}$ e-cm.
Furthermore, the works in \cite{Barbieri2,Okada,Babu2} are related
to analysis of the EDM of leptons in supersymmetric models . In
the work \cite{Babu}, the lepton EDMs are studied by scaling them
with corresponding lepton masses and the EDM of electron is
predicted as $10^{-27}$ e-cm. \cite{Ilt1} is devoted to the study
of the EDM of leptons $e,\mu,\tau$ in the general two Higgs
doublet model (2HDM). Lepton EDM in non-degenerate supersymmetric
see saw model have been examined in \cite{Ellis} and it was shown
that in the minimal supersymmetric seesaw model with
non-degenerate heavy neutrino masses, the charged lepton EDMs may
be enhanced by several orders of the magnitude compared to the
heavy neutrino case.  In \cite{Abel} it has been emphasized that
there was another source of EDM inherent in more fundamental
models such as supersymmetric models derived from strings and the
string models which accommodate the fermion mass hierarchy and
mixings generally lead to large EDMs. The EDM of leptons have been
studied in the left right supersymmetric models in \cite{Frank}.
\cite{Kong} is devoted to the sources of fermion dipole moment
contributions from R-parity violating or lepton number violating
parameters. In \cite{Pilaftsis} the muon electric dipole moment
induced by Higgs bosons, third-generation quarks and squarks,
charginos and gluinos in the minimal supersymmetric standard model
(MSSM) has been studied. The constraints that the non-observation
of electric dipole moments imposed on the radiatively-generated
CP-violating Higgs sector and on the mechanism of electroweak
baryogenesis in the MSSM, were discussed. The tau EDM has been
examined in \cite{Bernabeu} and it was concluded that the
stringent and independent bounds on the tau EDM competitive with
the high energy measurements, can be established in the low
energies experiments.

The naturalness bounds on the dipole moments from new physics has
been studied on \cite{Akama}. The EDMs of quarks are not directly
measurable, however they can affect the hadron phenomenology, for
example through the escaling violation in deep inelastic
lepton-hadron scattering or EDM of nucleons. In this work the
upper bounds of the quark EDMs have been estimated as
\begin{eqnarray}
|d^\gamma_{u}|&\leq& 4.0\times 10^{-20}\, e-cm \nonumber \, , \\
|d^\gamma_{d}|&\leq& 1.5\times 10^{-19}\, e-cm\nonumber \, , \\
|d^\gamma_{s}|&\leq& 3.0\times 10^{-18}\, e-cm\nonumber \, , \\
|d^\gamma_{c}|&\leq& 1.1\times 10^{-17}\, e-cm\nonumber \, , \\
|d^\gamma_{b}|&\leq& 7.0\times 10^{-17}\, e-cm\nonumber \, , \\
|d^\gamma_{t}|&\leq& 1.5\times 10^{-15}\, e-cm \,  .
\label{EDMeq1}
\end{eqnarray}
In \cite{Gomez}, the electric dipole form factors for heavy
fermions have been calculated in the 2HDM and it was predicted as
\begin{eqnarray}
|d^\gamma_{\tau}| & \leq& 10^{-23}\, e-cm\nonumber \, , \\
|d^\gamma_{t}| & \leq& 10^{-20}\, e-cm\nonumber \, , \\
|d^\gamma_{b}| & \leq& 10^{-20}\, e-cm\nonumber \, , \\
|d^\gamma_{c}| & \leq&10^{-21}\, e-cm .\label{EDMeq2}
\end{eqnarray}
where the predictions for  model I and II versions of the 2HDM are
\begin{eqnarray}
|d^\gamma_{\tau}| & \leq& 10^{-24}\, e-cm\nonumber \, , \\
|d^\gamma_{t}|& \leq& 10^{-20}\, e-cm\nonumber \, , \\
|d^\gamma_{b}|& \leq& 10^{-20}e-cm\nonumber \, , \\
|d^\gamma_{c}|& \leq&10^{-24}\, e-cm \, . \label{EDMeq3}
\end{eqnarray}

The EDM of $b$-quark in the general 2HDM  and in the 3HDM with
$O(2)$ symmetry in the Higgs sector has been calculated in
\cite{Ilt2} and, even at one loop, the EDM was obtained  in the
order of $10^{-20}$e-cm. The work in \cite{Ilt3} is devoted to the
EDM of top quark in the general 2HDM, including charged Higgs
contribution and the numerical value of EDM was obtained at the
order of magnitude of $10^{-20}$ e-cm.

The WEDM of an elementary particle is the another signature for
the CP violation. An improved test of CP invariance in the
reaction $e^+ e^-\rightarrow \tau^+ \tau^-$ on $Z^0$ peak is
performed using the data sample recorded between $1991$ and $1995$
with the OPAL detector at LEP. From the non-observation of CP
violation, the upper limits for the real and imaginary part of the
WEDM of tau lepton is derived $95 \%$ confidence level as
$Re[d^Z_{\tau}]\leq 5.6 \times 10^{-18}$ e-cm and
$Im[d^Z_{\tau}]\leq 1.5\times 10^{-17}$ e-cm \cite{Opal}.

In \cite {Bernabeu2} the WEDMs of fermions have been analyzed and
it was emphasized their property of being five-dimensional
operators in the effective Lagrangian. This suggest that WEDM is
proportional to $m_f/\Lambda^2$ where $\Lambda$ is the scale of
the physics involved. A theoretical work was performed in the
framework of the leptoquark models \cite{Bernreuther} and the WEDM
of tau is predicted as $|d^Z_{\tau}|\sim 10^{-18}$ e-cm. In
\cite{Bernabeu3} the WEDM of $\tau$ lepton has been estimated in
the order of $10^{-18}$ e-cm by studying the normal and transverse
polarization for $\tau^+ \tau^-$ pairs produced from unpolarized
$e^+ e^-$ collisions at the Z-peak.

In \cite{Gomez} the WEDM of heavy fermions have been studied in
the framework of the 2HDM and the predicted values read:
\begin{eqnarray}
|d^Z_{\tau}|& \leq& 10^{-22}\, e-cm\nonumber \, , \\
|d^Z_{t}|& \leq& 10^{-19}\, e-cm\nonumber \, , \\
|d^Z_{b}|& \leq& 10^{-20}\, e-cm\nonumber \, , \\
|d^Z_{c}|& \leq&10^{-22}\, e-cm \label{dZ1} \, .
\end{eqnarray}
These numerical results agree with the predictions of $\tau$
lepton and $b$ quark WEDMs which were obtained in \cite{Hollik}
within the MSSM.

The AMM of a fermion receives its leading value from one-loop
corrections in the SM and therefore it is not so much sensitive to
the new physics effects beyond. However, with the  recent
anouncement of world average AMM of muon at BNL \cite{BNL},
\begin{eqnarray}
a_{\mu}=11\, 659\, 203\, (8)\times 10^{-10} \label{AMMa}\,\, ,
\end{eqnarray}
which has about half of the uncertainty of previous measurements
and its SM prediction, a new window is opened for testing the SM
and beyond, assuming that the new physics effects can not exceed
the deviation. There are various attempts in different models to
explain the small deviation of the SM result \cite{Lane}.
Furthermore the deviation of  AMM of electron (tau) from its SM
contribution is $\Delta a_{e}\, (\Delta a_{\tau}) \sim 10^{-11}\,
(10^{-3})$ \cite{Akama} and this can be explained by the physics
beyond the SM. In \cite{Akama} AMM of quarks have been also
studied and they have been estimated as:
\begin{eqnarray}
|a^\gamma_{u}|&\leq& 4\times 10^{-6}\mu_N \nonumber \, , \\
|a^\gamma_{d}|&\leq& 1.5\times 10^{-5} \mu_N\nonumber \, , \\
|a^\gamma_{s}|&\leq& 3\times 10^{-4}\mu_N\nonumber \, , \\
|a^\gamma_{c}|&\leq& 1.1\times 10^{-3}\mu_N\nonumber \, , \\
|a^\gamma_{b}|&\leq& 7 \times10^{-3}\mu_N \nonumber \, , \\
|a^\gamma_{t}|&\leq& 1.4\times 10^{-1}\mu_N \nonumber \, ,
\end{eqnarray}
where $\mu_N=\frac{e}{2\,m_{proton}}$ is the nuclear magneton.

The WAMM of a fermion receives its leading value from one-loop
corrections in the SM similar to AMM. Since WAMM is generated
through a chirality flip mechanism it is expected to be
proportional to the mass of the particle and therefore, the heavy
fermions are the candidates for the sizable WAMMs. WAMM of $\tau$
lepton has been studied in the literature extensively
\cite{Sanchez}-\cite{ALEPH}. L3 presented  the direct limit on
$|Re(a^Z_{\tau})| \leq 0.014 95$ $\%$ CL \cite{Sanchez}. Using the
data given in 1997 \cite{Bohm} the WAMM has been predicted as
$|Re(a^Z_{\tau})| \leq 0.0027 95$ $ \%$ CL \cite{Rizzo}.

This physical quantity has been calculated in the framework of the
SM as
\begin{eqnarray}
a^Z_{\tau}(SM)=-(2.10+0.61 i)\times 10^{-6} \, ,
\end{eqnarray}
in \cite{Bernabeu4} and it was studied in \cite {Comelli} in the
framework of the 2HDM. The work in \cite{Carlos} is devoted to the
calculation of the WAMM in the super symmetric (SUSY) model and it
was observed that the WAMM increases with the increasing values of
the parameter $tan\,\beta$. The WAMM of $b$-quark has been
calculated as
\begin{eqnarray}
a^Z_{b}=(3.57-1.95 i)\times 10^{-4}  \, ,
\end{eqnarray}
in \cite{Comelli} and  in \cite{Bernabeu5} the numerical values
for $c$ and $b$ quark were presented as
\begin{eqnarray}
a^Z_{c}&=&(-2.80+1.09 i)\times 10^{-5} \nonumber \, , \\
a^Z_{b}&=&(2.98-1.56 i)\times 10^{-4} \, .
\end{eqnarray}
Notice that, similar to the WEDM, WAMM is created by
five-dimensional operators in the effective Lagrangian and
therefore WAMM is proportional to $m_f^2/\Lambda^2$ where
$\Lambda$ is the scale of the physics involved \cite{BNL}.

In the present work, we study the noncommutative (NC) effects on
the dipole and magnetic moments of fermions in the framework of
the SM. The string theory arguments re-motivate the physics on the
noncommutative spaces \cite{Connes2,Witten} and the quantum field
theory over noncommutative spaces \cite{Connes} has been reached a
great interest in recent years .

NC field theory have a non-local structure and the Lorentz
symmetry is explicitly violated. These violations have been
analyzed in \cite{Mocioiu,Carlson1}. The renormalizability and the
unitarity of NC theories have been studied in the series of works
\cite{Gonzales}, \cite{Gomis} and \cite{Hewett}. The quantum
electrodynamics including the noncommutative effects (NCQED) has
been examined in \cite{Riad,Hayakawa} and NCQED has been studied
in the extra dimensions in \cite{Carlson2}. The noncommutativity
in non-abelian case has been formulated in
\cite{Krajewski,Madore}. The work in \cite{Calmet} is due to the
application of this formulation in to the SM. Noncommutative SM
(NCSM) building has been studied in \cite{Chaichian} and the
determination of triple neutral gauge boson couplings has been
done in \cite{Deshpande}. In \cite{Xiao}  a unique model for
strong and electroweak interactions with their unification has
been constructed. Furthermore, the phenomenological analysis of
the noncommutative effects on some processes has been studied in
several works \cite{Hinchliffe}-\cite{IltNon3}. In
\cite{Hinchliffe}, the noncommutative CP violating effects have
been examined at low energies and it was ephasized that CP
violation due to noncommutative geometry was comparable to the one
due to the standard model (SM) only for a noncommutative scale
$\Lambda \leq 2\, TeV$. \cite{Behr} is devoted to the SM forbidden
processes $Z\rightarrow \gamma\gamma$ and $Z\rightarrow gg$ with
the inclusion of the NC effects. In \cite{Iltanbsgl} the form
factors, appearing in the inclusive $b\rightarrow s g$ decay, has
been calculated in the NCSM, using the approximate phenomenology
and the new operators existing in $b\rightarrow s g$ decay due to
the NC effects has been obtained in \cite{Iltanbsgam}. In the
recent work \cite{Cang}, the possible effects of NC geometry on
weak CP violation and the untarity triangles have been examined.
The work in \cite{IltNon3} is devoted to the $Z\rightarrow l^+
l^-$ and $W\rightarrow \nu_{l} l^+$ decays, for $l=e,\mu,\tau$, in
the SM, including the noncommutative (NC) effects.

In our analysis we observe that the EDM and WEDM of fermions are
sensitive to the NC effects and these physical quantities would be
informative in the determination of the upper bounds of the NC
parameters.

The paper is organized as follows: In Section 2, we present the
explicit expressions for the EDM, AMM, WEDM and WAMM of fermions
in the framework of the NC SM. Section 3 is devoted to discussion
and our conclusions.
\section{The noncommutative effects on the dipole moments of the fermions
in the SM}
The nature of the space-time changes at very short distances of
the order of the Planck length and the non commutativity in the
space-time is a possible candidate to describe the physics at
these distances. In the noncommutative geometry the space-time
coordinates are replaced by Hermitian operators $\hat{x}_{\mu}$
which satisfy the equation \cite{Synder}
\begin{eqnarray}
[\hat{x}_{\mu},\hat{x}_{\nu}]=i\,\theta_{\mu\nu} \, , \label{com1}
\end{eqnarray}
where $\theta_{\mu\nu}$ is a real and antisymmetric tensor with
the dimensions of [$mass^{-2}$]. Here $\theta_{\mu\nu}$ can be
treated as a background field relative to which directions in
space-time is distinguished and its components are assumed as
constants over cosmological scales. Introducing $*$ product of
functions, instead of the ordinary one,

\begin{eqnarray}
(f*g)(x)=e^{\frac{i}{2}\,\theta_{\mu\nu} \,\partial^y_{\mu}\,
\partial^z_{\nu}} f(y)\, g(z)|_{y=z=x}\,.
\label{product}
\end{eqnarray}
it is possible to pass to the noncommutative field theory. The
commutation of the Hermitian operators $\hat{x}_{\mu}$ (see eq.
(\ref{com1})) holds with this new product, namely,
\begin{eqnarray}
[\hat{x}_{\mu},\hat{x}_{\nu}]_*=i\,\theta_{\mu\nu} \,\, .
\label{com2}
\end{eqnarray}

Since the noncommutative effects are tiny to observe in the low
energy physics one would expect that the physical quantities,
which do not exist even at the loop levels more than two in the
SM, may be sensitive to the NC effects. The EDM and WEDM can exits
at the higher order in the coupling constant in the SM since it
receives the non-vanishing value through the CP violating effects
coming from the CKM matrix elements. Therefore these quantities
may be sensitive to the new physics effects coming from the NC
extension of the SM. On the other hand, AMM and WAMM of a fermion
receive their leading value from one-loop corrections and
therefore it is not so much sensitive to the new physics effects
beyond the SM. However, with the stringent bounds obtained in the
measurements, the deviations of these quantities from their SM
values would be a possible candidate to search the NC effects. In
the present work we estimated the EDM, WEDM, AMM and WAMM of
massive leptons and quarks in the NC extension of the SM.

EDM (WEDM) and AMM (WAMM) exist in the case of the
photon-fermion-fermion (Z boson-fermion-fermion) coupling and the
effective Lagrangian of these quantities are
\begin{eqnarray}
{\cal L}_{EDM\, (WEDM)}=i d^{\gamma (Z)} \,\bar{f}\,\gamma_5
\,\sigma_{\mu\nu}\,f\, F^{\mu\nu} \,\, , \label{EDM1}
\end{eqnarray}
and
\begin{eqnarray}
{\cal L}_{AMM\, (WAMM)}=a^{\gamma (Z)} \, \frac{e
(g)}{4\,m_f}\,\bar{f} \,\sigma_{\mu\nu}\,l\, F^{\mu\nu} \,\, ,
\label{AMM1}
\end{eqnarray}
where $F_{\mu\nu}$ is the electromagnetic field tensor (weak field
tensor), $d^{\gamma (Z)}$ is EDM (WEDM) of the fermion $f$ and
$a^{\gamma (Z)}$ is AMM (WAMM) of the fermion $f$.
Notice that EDM (WEDM) and AMM (WAMM) are proportional to the
coefficients $d$ and $a$ in the interactions
\begin{eqnarray}
& & d\, \hat{n}. \vec{E} \nonumber \, ,\\
& & \frac{e (g)}{4\, m_f}\, a \,\hat{n}. \vec{B}\, ,
\label{EDMAMM1}
\end{eqnarray}
respectively, as it can be obtained using the eqs. (\ref{EDM1})
and (\ref{AMM1}).

When the non-commutative effects are switched on there exists a
new contribution which is proportional to the a function of the
noncommutative parameter $\theta$. The Lagrangian for the
additional vertex to the $\gamma f^+ f^-$ and $Z f^+ f^-$
interaction reads \cite{Calmet}
\begin{eqnarray}
L^f_{A (Z)}&=&-\frac{i}{2}\,\Bigg\{ c^{A (Z)}_{1\,f} \bar{f}
\Bigg(\frac{1}{2}\theta_{\mu\nu}
\gamma_{\alpha}+\theta_{\nu\alpha} \gamma_{\mu} \Bigg)\,L\,
\partial^{\alpha} f + c^{A (Z)}_{2\,f} \bar{f} \Bigg(
\frac{1}{2}\theta_{\mu\nu} \gamma_{\alpha}+\theta_{\nu\alpha}
\gamma_{\mu} \Bigg)\, R\, \partial^{\alpha} f \Bigg\} \nonumber \\
& & (\partial^{\mu} V_{A (Z)}^{\nu}- \partial^{\nu} V_{A
(Z)}^{\mu}) \,\, , \label{LagrangNC}
\end{eqnarray}
with
\begin{eqnarray}
c^{A}_{1\,f}&=&cos\theta_W g' Y^L_f+\frac{1}{2}\,s\, g \,sin
\theta_W\,
, \nonumber \\
c^{A}_{2\,f}&=&cos\theta_W \,g' \, Y^R_f\, , \nonumber \\
c^{Z}_{1\,f}&=&-sin\theta_W \,g' \,Y^L_f+\frac{1}{2}\,s \,g\,
cos\theta_W\,, \nonumber \\
c^{Z}_{2\,f}&=&-sin\theta_W \,g' \,Y^R_f \, ,  \label{c12}
\end{eqnarray}
where $V_A^{\mu}$ ($V_Z^{\mu}$) denotes the photon (Z boson)
field, $s=1\,(-1)$ for $u$ quarks ($d$ quarks and leptons),
$Y^L_f=(-\frac{1}{2}, \frac{1}{6})$ for ($f$=leptons, $f$=quarks)
and $Y^R_f=(-1, \frac{2}{3}, \frac{-1}{3})$ for ($f$=leptons,
$f$=up quarks, $f$=down quarks).

At this stage we formulate the dipole moments of fermions using
their definitions in eqs. (\ref{EDMAMM1}) and the NC extensions of
SM given in eq. (\ref{LagrangNC}). The EDM (WEDM) interaction of
fermions are obtained as
\begin{eqnarray}
L^{f\,NC}_{EDM\, (WEDM)}= \frac{1}{2}\,e\, c_f\, m_f\,
|\Theta_T|\,\hat{p}_i. E^i \, , \label{LEDM}
\end{eqnarray}
where we take $\bar{f}\theta_{0i}f=|\Theta_T|\,\hat{p}_i$ and we
use $F^{0i}=E^i$. In eq. (\ref{LEDM}) the vector $(\Theta_T)_i$ is
responsible for time-space non commutativity and $\hat{p}_i$ is
the unit vector in the direction of $(\Theta_T)_i$. Furthermore,
$E_i$ is the electric (weak electric) field and $m_f$ is the
fermion mass. Finally the EDM (WEDM) of fermions reads
\begin{eqnarray}
d_{f}&=& \frac{1}{2}\,e\, c_f\, m_f\, |\Theta_T| \, ,
\label{LQEDM}
\end{eqnarray}
where $c_f=Q_f$ for the EDM case and
$c_f=\frac{1}{6\,sin2\theta_W}\, (3-8 sin^2\theta_W)$,
$-\frac{1}{6\,sin 2\theta_W}\, (3-4 sin^2\theta_W)$,
$-\frac{1}{2\,sin 2\theta_W}\, (1-4 sin^2\theta_W)$ for $u$
quarks, $d$ quarks and massive leptons respectively, in the case
of WEDM.

The AMM (WAMM) can be obtained using the interaction lagrangian
\begin{eqnarray}
L^{NC}_{f\, AMM \,(WAMM)}&=& \frac{1}{2}\, b_f\, m^2_f\,
|\Theta_S|\,\hat{\mu}. \vec{B} \, , \label{LAMM}
\end{eqnarray}
where we take
$\bar{f}\theta_{ij}f=\frac{1}{2}\,\epsilon_{ijk}\,|\Theta_S|\,
\hat{\mu}^k$ and we use $\epsilon_{ijk}\,F^{ij}=B_k$. Here $B_k$
is the magnetic (weak magnetic) field . The vector $(\Theta_S)_k$
is responsible for space-space non commutativity and $\hat{\mu}^k$
is the unit vector in the direction of $(\Theta_S)^k$. As a
result, the AMM (WAMM) of fermions reads
\begin{eqnarray}
a_{f}= \frac{1}{2}\, b_f\, m^2_f\, |\Theta_S| \, , \label{LQAMM}
\end{eqnarray}
where $b_f=Q_f$ for the AMM case and $b_f=\frac{1}{6\,sin
2\theta_W}\,(3-8 sin^2\theta_W)$, $-\frac{1}{6\,sin 2\theta_W}\,
(3-4 sin^2\theta_W)$, $-\frac{1}{2\,sin 2\theta_W}\, (1-4
sin^2\theta_W)$ for $u$ quarks, $d$ quarks and massive leptons
respectively, in the case of WAMM.
The eqs (\ref{LQEDM}) and (\ref{LQAMM}) show that the NC effects
on the EDM and WEDM (AMM and WAMM) are proportional to the mass
(square of mass) of the fermions and the heavy fermion EDM and
WEDM (AMM and WAMM) receive the large contribution from the NC
effects. Notice that the AMM and WAMM is due to the intrinsic
magnetic moment of the fermion and its is spin independent.
\section{Discussion}
This section is devoted to the analysis of the NC effects on the
dipole moments of fermions EDM, WEDM, AMM and WAMM,  in the
framework of the SM. Since the EDM and WEDM can exits at the
higher order in the coupling constant in the SM due to the their
CP violating nature, they could be sensitive to the new physics
effects beyond, NC effects in the present work. The NC effects on
the dipole moments deserve to analyze since they can bring
comprehensive information in the determination of the bounds of
the new NC parameters. On the other hand, AMM and WAMM of a
fermion exist in the one-loop order in the SM, and  they are not
sensitive to the new physics effects beyond, compared to the EDM
and WEDM. However, the deviations of the these physical quantities
from the SM values can be examined by including the NC effects and
the possible constraints can be obtained for the NC parameters.

Our starting point is the experimental result of the electron EDM
$|d_e|\leq 3.4\times 10^{-26}\, e-cm$ \cite{Hudson}. Here we
assume that the source of the EDM is the NC effect in the SM.
Using the experimental upper limit of $|d_e|$ and the eq.
(\ref{LQEDM}) we obtain an upper bound for the NC parameter
$|\Theta_T|$ which is responsible for time-space non
commutativity, $|\Theta_T|\leq 6.67\times 10^{-10} \,(GeV^{-2})$.
Now we estimate the upper limits of the EDM of the fermions
(leptons $\mu$, $\tau$ and quarks) using the numerical value of
$|\Theta_T|$ and the eq (\ref{LQEDM}). Notice that in our
predictions we study the absolute values of the dipole moments.

\begin{figure}[htb]
\vskip -3.0truein \centering \epsfxsize=6.8in
\leavevmode\epsffile{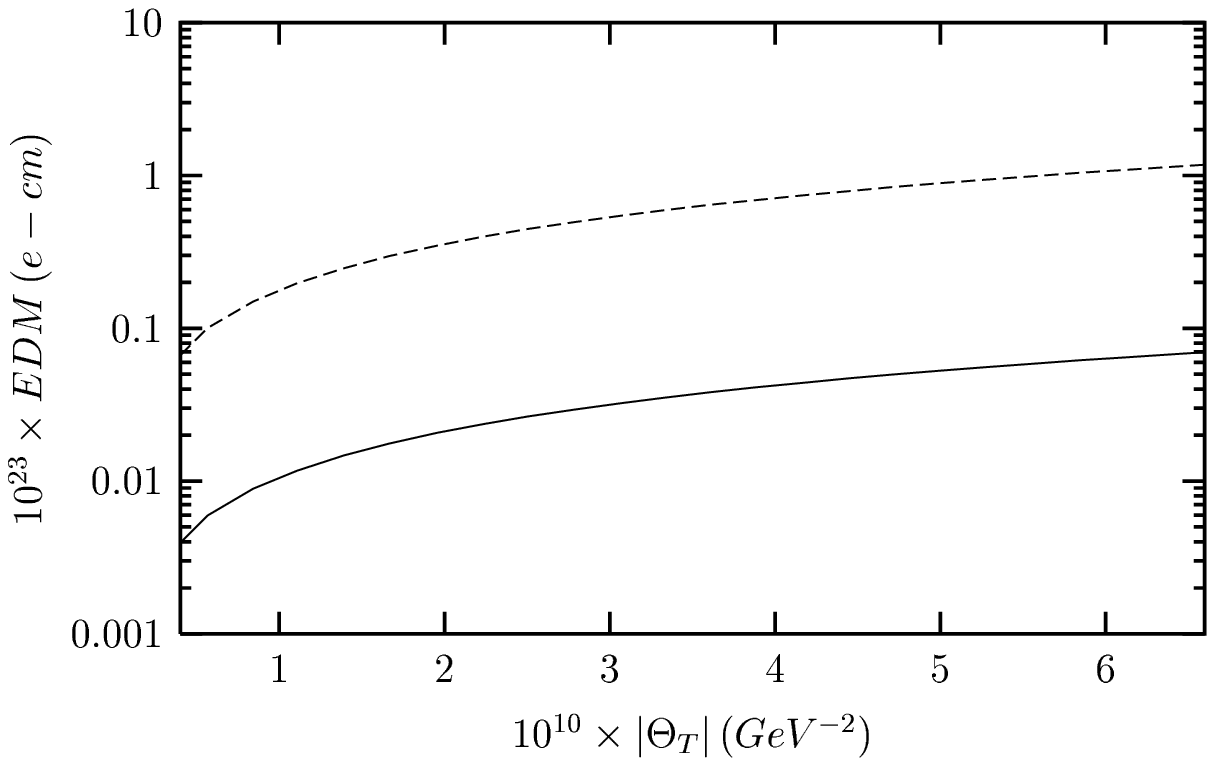} \vskip -3.0truein \caption[]{
The noncommutative parameter $|\Theta_T|$ dependence of lepton
EDM. Here the solid (dashed) line represents EDM of $\mu$ ($\tau$)
lepton.} \label{EDMLepton}
\end{figure}

In Fig. \ref{EDMLepton}, we present the noncommutative parameter
$|\Theta_T|$ dependence of lepton EDM.  Here the solid (dashed)
line represents EDM of $\mu$ ($\tau$) lepton. This figure shows
that $\tau$ ($\mu$) lepton EDM can take the numerical values at
the order of  $10^{-23}$ e-cm ($10^{-24}$ e-cm) in the given
interval of the parameter $|\Theta_T|$ . These numerical values
satisf the current experimental results, $d_{\mu} =(3.7\pm
3.4)\times 10^{-19}$ \cite{Bailey} and $d_{\tau} \leq
3.1\times10^{-16}$ \cite{Groom}.

\begin{figure}[htb]
\vskip -3.0truein \centering \epsfxsize=6.8in
\leavevmode\epsffile{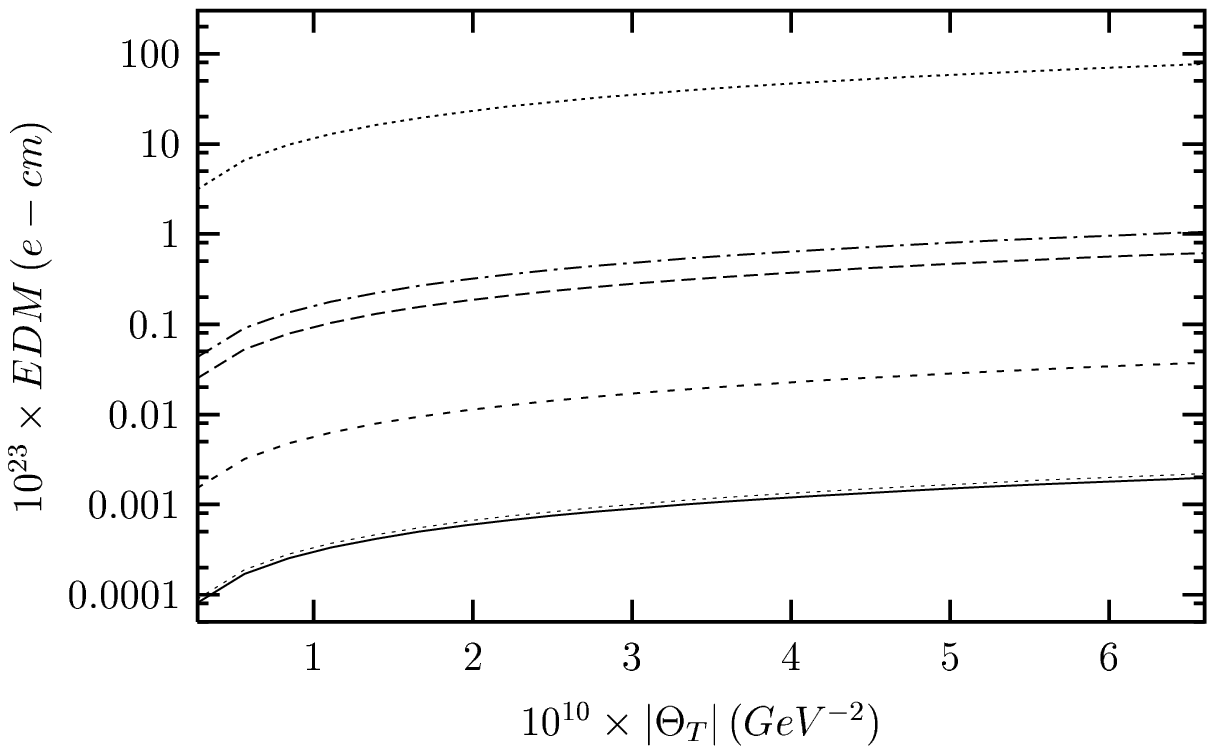} \vskip -3.0truein \caption[]{The
noncommutative parameter $|\Theta_T|$ dependence of quark EDM.
Here the solid (dotted, small dashed, dashed, dashed-dotted,
dense-dotted) line represents EDM of $u$ ($d$, $s$, $c$, $b$, $t$)
quark.} \label{EDMQuark}
\end{figure}

Fig. \ref{EDMQuark} is devoted to the noncommutative parameter
$|\Theta_T|$ dependence of quark EDM.  Here the dotted (solid,
small dashed, dashed, dashed-dotted, dense-dotted) line represents
EDM of $u$ ($d$, $s$, $c$, $b$, $t$) quark. $u$ ($d$, $s$, $c$,
$b$, $t$) EDM can receive the numerical values at the order of
magnitude of $10^{-26}$ e-cm ($10^{-26}, 5.0\times 10^{-25},
10^{-23}, 10^{-23}, 10^{-21}$ e-cm) for $u$ ($d$, $s$, $c$, $b$,
$t$) quark. These numerical values are in accordance with the
theoretical results presented in eqs. (\ref{EDMeq1}, \ref{EDMeq2},
\ref{EDMeq3}) and the t-quark EDM which is calculated in
\cite{Ilt3}.

\begin{figure}[htb]
\vskip -3.0truein \centering \epsfxsize=6.8in
\leavevmode\epsffile{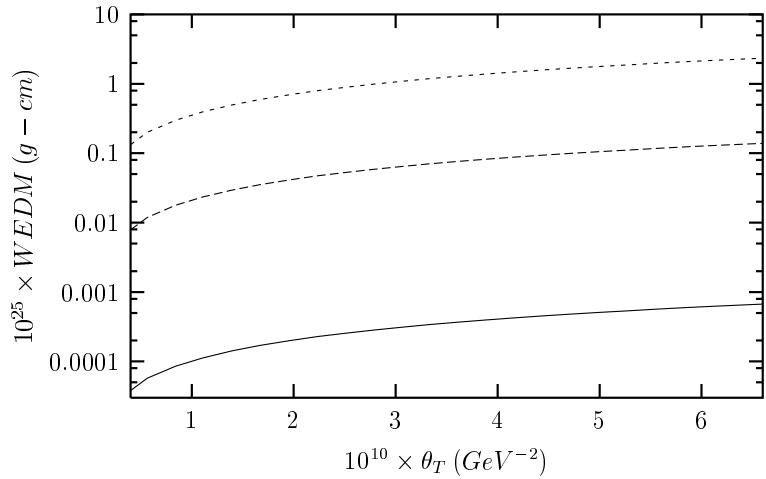} \vskip -3.0truein \caption[]{
The noncommutative parameter $|\Theta_T|$ dependence of lepton
WEDM. Here the solid (dashed, small dashed) line represents WEDM
of $e$ ($\mu$, $\tau$) lepton.} \label{WEDMLepton}
\end{figure}

Fig. \ref{WEDMLepton}, represents the noncommutative parameter
$|\Theta_T|$ dependence of lepton WEDM. Here the solid (dashed,
small dashed) line represents WEDM of $e$ ($\mu$, $\tau$) lepton.
The $\tau$ lepton WEDM can take the numerical values at the order
of the magnitude of $10^{-25}\, g-cm$. This results is almost
eight order smaller compared to the experimental result
$Re[d^Z_{\tau}]=\leq 5.6 \times 10^{-18}$ e-cm and
$Im[d^Z_{\tau}]=\leq 1.5\times 10^{-17}$ e-cm \cite{Opal}. The
more accurate measurements in future would make it possible to
test NC effects on WEDM more stringently.  Furthermore, the
theoretical result in the framework of the 2HDM was presented in
\cite{Gomez} as $d^Z_{\tau} \leq 10^{-22}$ e-cm and the $\tau$
lepton WEDM coming from the NC effects is almost three orders
smaller than this numerical value, in the given NC parameter
region. The electron and $\mu$ lepton WEDM can receive to the
numerical values at the order of the magnitude of $10^{-28}$ g-cm
and $10^{-26}$ g-cm and they can be tested with more sensitive
forthcoming experimental measurements.

\begin{figure}[htb]
\vskip -3.0truein \centering \epsfxsize=6.8in
\leavevmode\epsffile{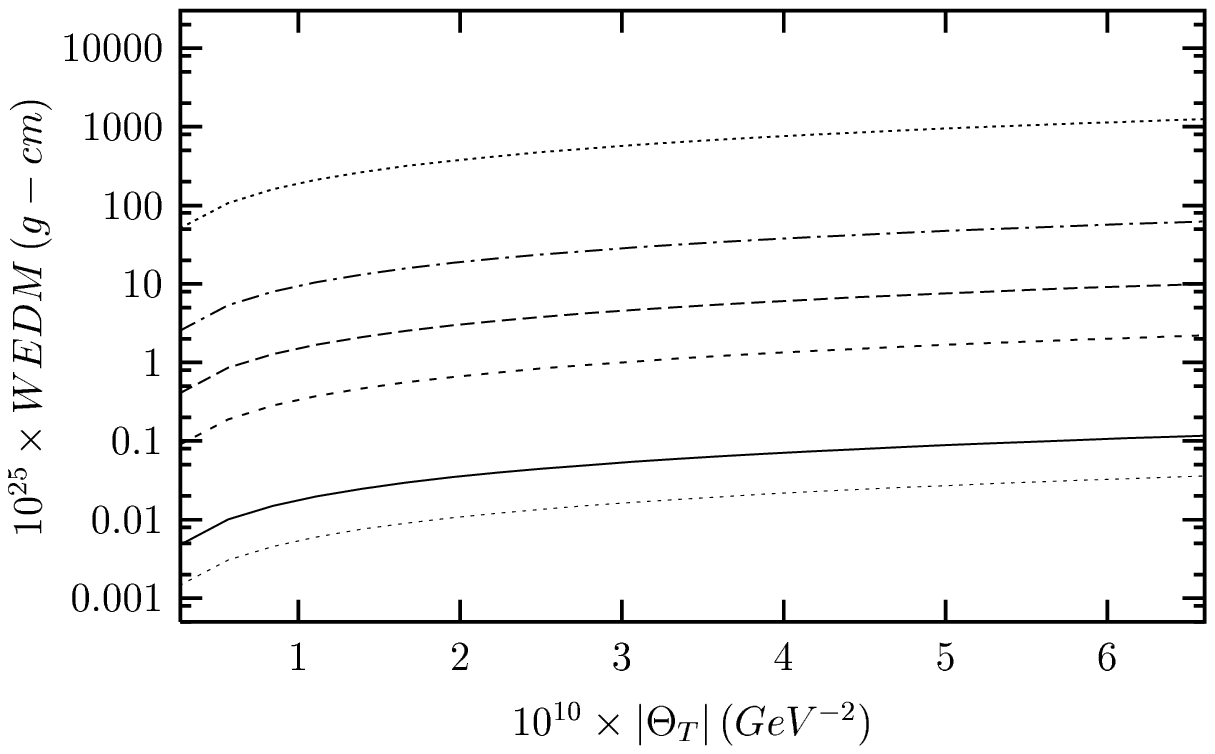} \vskip -3.0truein
\caption[]{The noncommutative parameter $|\Theta_T|$ dependence of
quark WEDM.  Here the solid (dotted, small dashed, dashed,
dashed-dotted, dense-dotted) line represents EDM of $u$ ($d$, $s$,
$c$, $b$, $t$) quark.} \label{WEDMQuark}
\end{figure}

In Fig. \ref{WEDMQuark} we show the noncommutative parameter
$|\Theta_T|$ dependence of quark WEDM.  Here the solid (dotted,
small dashed, dashed, dashed-dotted, dense-dotted) line represents
WEDM of $u$ ($d$, $s$, $c$, $b$, $t$) quark. It is observed that
the  WEDM can take the numerical value at the order of magnitude
of $5.0\times 10^{-27}$ e-cm ($10^{-26}, 10^{-25}, 10^{-24},
10^{-23}, 10^{-22}$ e-cm) for $u$ ($d$, $s$, $c$, $b$, $t$) quark.
These upper bounds are smaller compared to the ones obtained in
\cite{Gomez} (see eq. \ref{dZ1}).

Now, we analyze the AMM and WAMM of a fermions in the SM including
the NC effects. These physical quantities can exist even in the
one-loop level in the SM and therefore they are not so much
sensitive to the new physics effects beyond. However, with the
accurate analysis of the deviations of the these physical
quantities from the SM values it would be possible to test the new
effects coming from NC geometry.

The recent anouncement of world average AMM of muon at BNL
\cite{BNL} in eq. (\ref{AMMa}) and the SM prediction forces that
there is still a standard deviation from the experimental result
and this could possibly be due to the effects of new physics, at
present. Now, we respect the assumption that the new physics
effects, NC effects in the present work, on the numerical value of
muon AMM should not exceed the present experimental uncertainty,
$\sim 10^{-9}$. Using the eq. (\ref{LAMM}), we obtain an upper
bound for the NC parameter $|\Theta_S|$ which is responsible for
space-space non commutativity, $|\Theta_S|\leq 1.68\times
10^{-7}\, (GeV^{-2})$ and  using this upper bound, we estimate the
NC effects on the  AMM of the fermions (leptons $e$, $\tau$ and
quarks), which is denoted as $\Delta{a}$.

\begin{figure}[htb]
\vskip -3.0truein \centering \epsfxsize=6.8in
\leavevmode\epsffile{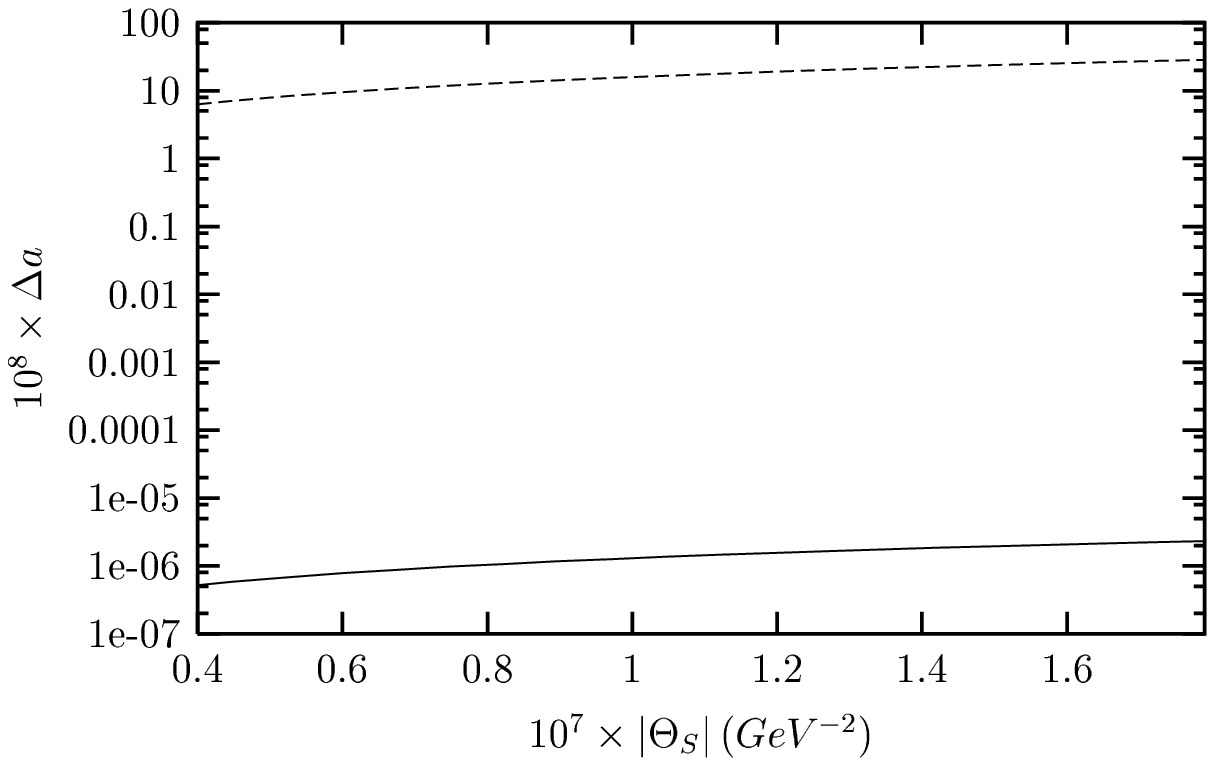} \vskip -3.0truein
\caption[]{The noncommutative parameter $|\Theta_S|$ dependence of
lepton $\Delta{a}$.  Here the solid (dashed) line represents
$\Delta{a}$ of $e$ ($\tau$) lepton.} \label{AMMLepton}
\end{figure}

In Fig. \ref{AMMLepton}, we present the noncommutative parameter
$|\Theta_S|$ dependence of lepton $\Delta{a}$.  Here the solid
(dashed) line represents $\Delta{a}$ of $e$ ($\tau$) lepton. This
figure shows that $\tau$ ($e$) lepton $\Delta{a}$ can take the
numerical values at the order of the magnitude of $5.0\times
10^{-7}$ ($10^{-14}$). These numerical values are far from the
results predicted in \cite{Akama} $\Delta a_{e}\sim 10^{-11}\, ,
(\Delta a_{\tau}\sim 10^{-3})$.

\begin{figure}[htb]
\vskip -3.0truein \centering \epsfxsize=6.8in
\leavevmode\epsffile{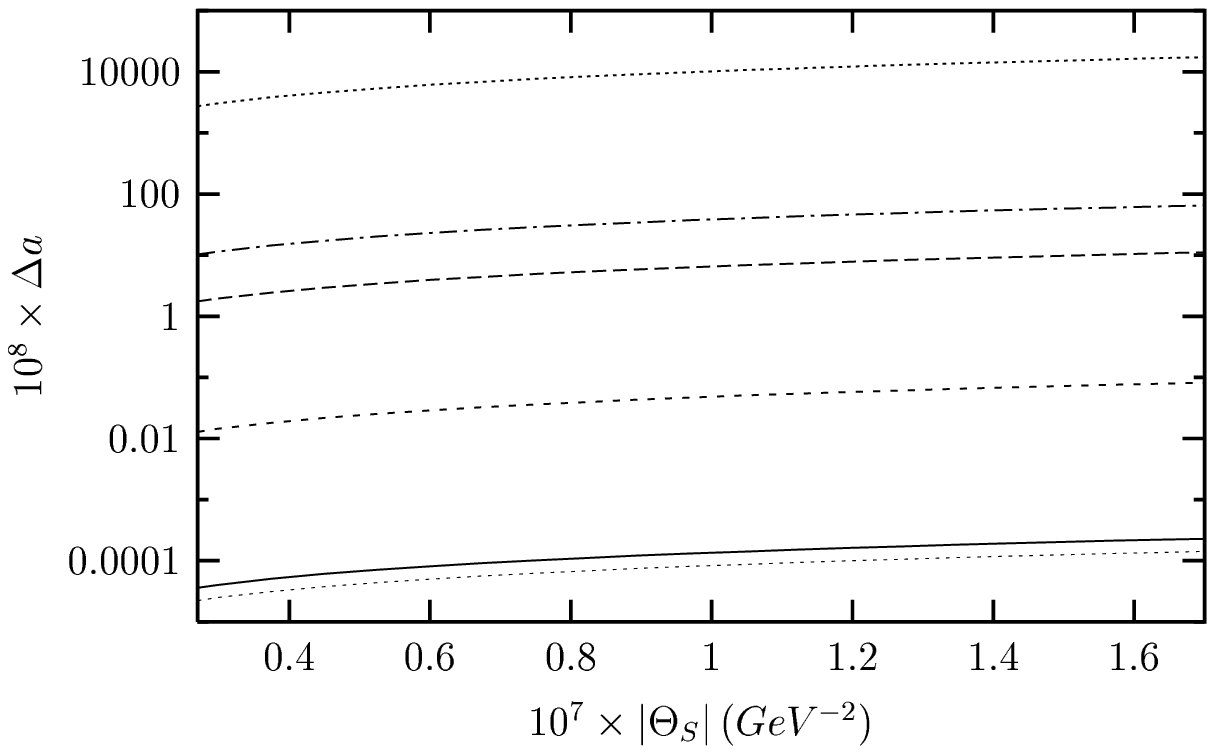} \vskip -3.0truein \caption[]{The
noncommutative parameter $|\Theta_S|$ dependence of quark
$\Delta{a}$.  Here the solid (dotted, small dashed, dashed,
dashed-dotted, dense-dotted) line represents $\Delta{a}$ of $u$
($d$, $s$, $c$, $b$, $t$) quark.} \label{AMMQuark}
\end{figure}

Fig. \ref{AMMQuark} is devoted to the noncommutative parameter
$|\Theta_S|$ dependence of quark $\Delta{a}$.  Here the solid
(dotted, small dashed, dashed, dashed-dotted, dense-dotted) line
represents $\Delta{a}$ of $u$ ($d$, $s$, $c$, $b$, $t$) quark.
$\Delta{a}$ can receive the numerical values in the order of
magnitude of $10^{-12}$ ($10^{-12}, 10^{-9}, 10^{-7}, 5\times
10^{-7}, 10^{-3}$) for $u$ ($d$, $s$, $c$, $b$, $t$) quark.

\begin{figure}[htb]
\vskip -3.0truein \centering \epsfxsize=6.8in
\leavevmode\epsffile{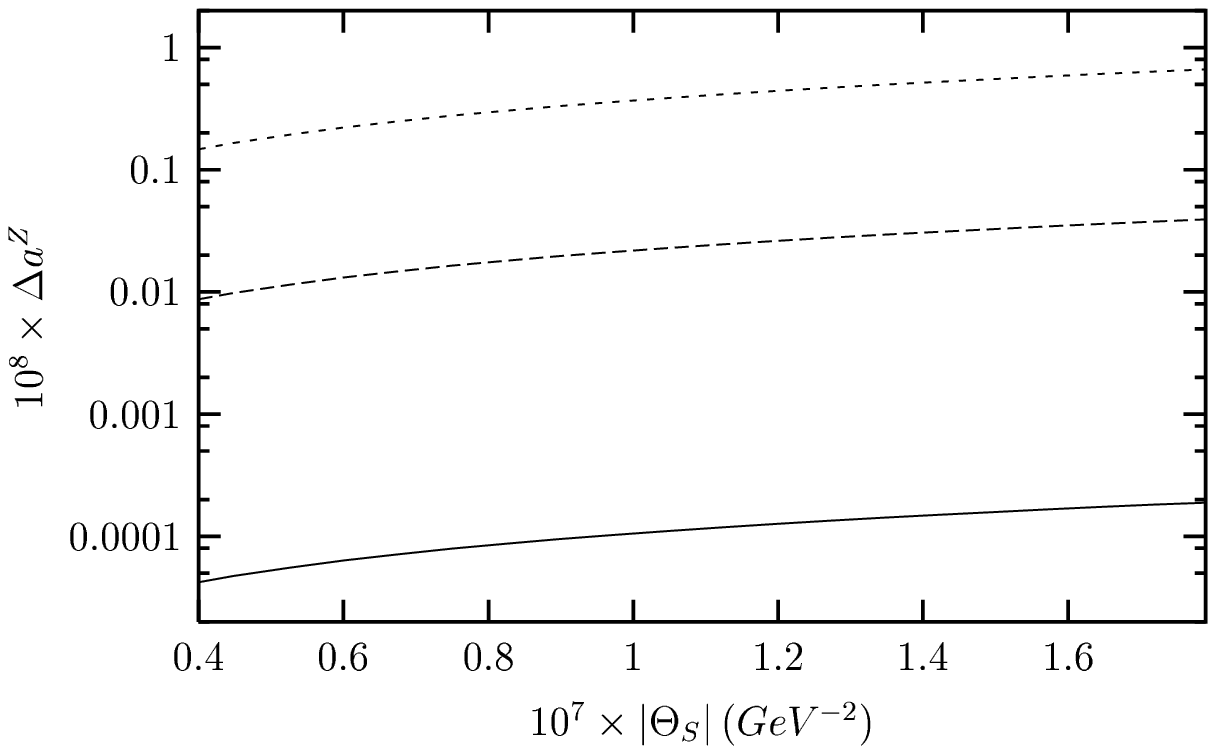} \vskip -3.0truein
\caption[]{The noncommutative parameter $|\Theta_S|$ dependence of
lepton WAMM, $\Delta{a}^Z$.  Here the solid (dashed, small dashed)
line represents $\Delta{a}^Z$ of $e$ ($\mu$, $\tau$) lepton.}
\label{WAMMLepton}
\end{figure}

Fig. \ref{WAMMLepton} represents the noncommutative parameter
$|\Theta_S|$ dependence of lepton WAMM, $\Delta{a}^Z$. Here the
solid (dashed, small dashed) line represents $\Delta{a}^Z$ of $e$
($\mu$, $\tau$) lepton. The NC effects on WAMM of $e$ ($\mu,
\,\tau$)lepton is at the order of the magnitude of $10^{-12}$
($5.0 \times 10^{-10}$, $10^{-8}$).

\begin{figure}[htb]
\vskip -3.0truein \centering \epsfxsize=6.8in
\leavevmode\epsffile{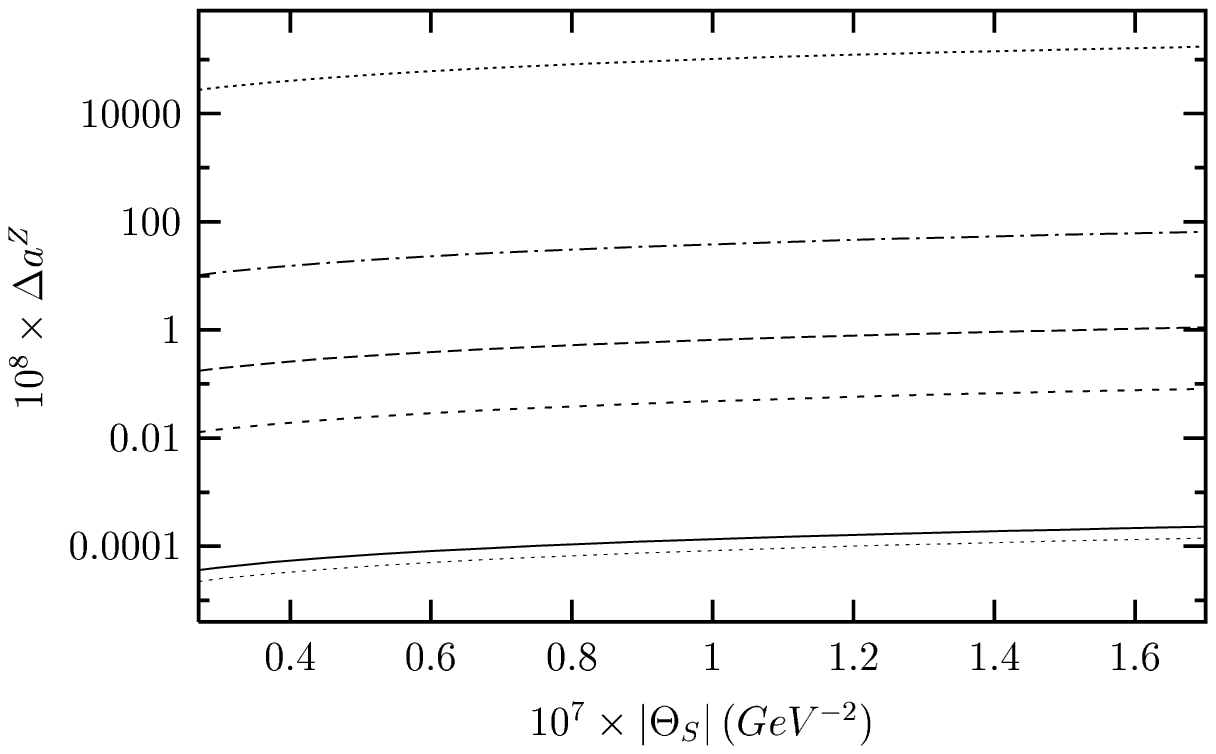} \vskip -3.0truein
\caption[]{The noncommutative parameter $|\Theta_S|$ dependence of
quark $\Delta{a}^Z$.  Here the solid (dotted, small dashed,
dashed, dashed-dotted, dense-dotted) line represents $\Delta{a}^Z$
of $u$ ($d$, $s$, $c$, $b$, $t$) quark} \label{WAMMQuark}
\end{figure}

Finally Fig. \ref{WAMMQuark} is devoted to the noncommutative
parameter $|\Theta_S|$ dependence of quark $\Delta{a}^Z$.  Here
the solid (dotted, small dashed, dashed, dashed-dotted,
dense-dotted) line represents $\Delta{a}^Z$ of $u$ ($d$, $s$, $c$,
$b$, $t$) quark. $u$ ($d$, $s$, $c$, $b$, $t$). $\Delta{a}$ can
receive the numerical values in the order of magnitude of
$10^{-12}$ ($10^{-12}, 10^{-9}, 10^{-8}, 10^{-6}, 10^{-4}$) for
$u$ ($d$, $s$, $c$, $b$, $t$) quark.

At this stage we would like to summarize our results:

\begin{itemize}
\item We take the experimental result of the electron EDM
$|d_e|\leq 3.4\times 10^{-26}$ e-cm \cite{Hudson} and obtain an
upper bound for the NC parameter $|\Theta_T|$ which is responsible
for time-space non commutativity, $|\Theta_T|\leq 6.67\times
10^{10} (GeV^{-2})$. Using this upper bound, we predict the EDM
and WEDM of the fermions existing with the NC effects. We observe
that the numerical values of EDM of the leptons $\mu$ and $\tau$
satisfy the current experimental results. The predicted quark EDMs
are in accordance with the theoretical results obtained in the
literature. Here the $\tau$ lepton WEDM can take the numerical
values at the order of  magnitude of $10^{-25}$ g-cm. This result
is almost eight order smaller compared to the experimental result
\cite{Opal} and almost three orders smaller compared to the
theoretical result given by  \cite{Gomez}. For the heavy quarks
$t$, $b$ and  $c$ the predicted values in the NC extension of the
SM is not so much far from the ones obtained in the framework of
the model I and II versions of the 2HDM \cite{Gomez}. With the
more accurate measurements in future, it would be possible to test
NC effects on WEDM more stringently.

\item We restrict the NC parameter $|\Theta_S|$ which is
responsible for space-space non commutativity as $|\Theta_S|\leq
1.68\times 10^{-7} (GeV^{-2})$ using the assumption that the NC
effects on the numerical value of muon AMM should not exceed the
present experimental uncertainty, $\sim 10^{-9}$. The numerical
values $\Delta{a}$ for $\tau$ and $e$ lepton we obtain are far
from the results predicted in the literature \cite{Akama}.

\item The NC effects on the EDM and WEDM (AMM and WAMM) are
proportional to the mass (square of mass) of the fermions and the
heavy fermion EDM and WEDM (AMM and WAMM) receive the large
contribution in the NCSM. Notice that the AMM and WAMM is due to
the intrinsic magnetic moment of the fermion and its is spin
independent.
\end{itemize}

Therefore, the more accurate experimental results of the dipole
moments of the fermions would be an effective tool to understand
the new physics effects  due to the NC geometry.
\section{Acknowledgement}
This work has been supported by the Turkish Academy of Sciences in
the framework of the Young Scientist Award Program.
(EOI-TUBA-GEBIP/2001-1-8)
%

%
\newpage
\end{document}